# Metric Expansion from Microscopic Dynamics in an Inhomogeneous Universe

Sascha Vongehr

Department of Philosophy and National Key Laboratory of Solid-State Microstructures, Department of Materials Science and Engineering, Nanjing University, Nanjing 210093, P. R. China

Theories with ingredients like the Higgs mechanism, gravitons, and inflaton fields rejuvenate the idea that relativistic kinematics is dynamically emergent. Eternal inflation treats the Hubble constant $H$ as depending on location. Microscopic dynamics implies that $H$ is over much smaller lengths than pocket universes to be understood as a local space reproduction rate. We illustrate this via discussing that even exponential inflation in TeV-gravity is slow on the relevant time scale. In our on small scales inhomogeneous cosmos, a reproduction rate $H$ depends on position. We therefore discuss Einstein-Straus vacuoles and a Lindquist-Wheeler like lattice to connect the local rate properly with the scaling of an expanding cosmos. Consistency allows $H$ to locally depend on Weyl curvature similar to vacuum polarization. We derive a proportionality constant known from Kepler's third law and discuss the implications for the finiteness of the cosmological constant.

## 1. Introduction and Overview

To model relativistic kinematics as an emergent property is ever more popular and the discussion draws an increasing number of contributions. Most point rightfully out that general relativity (GR) strongly discourages[1] the idea that objects are dynamically interacting with a background (medium). Nevertheless, the *relativistic kinematics from dynamics* concept is deeply rooted in modern physics. In contrast to loop quantum gravity, string theory has gravitons that interact and thereby "give rise" to gravity, at least in the sense of that the choice of background must be a consistent one. Rest mass in the standard model, i.e. even pure inertia against non-gravitational acceleration, is widely thought to be a permanently ongoing interaction with the Higgs field. The dynamics of inflaton fields is necessary in modeling the inflation of the early universe.

The viewpoint that GR could be an emergent property has a long history (reviewed thoroughly elsewhere[2,3]). It was already shown[4] in 1945 that a crystal-like Dirac-sea mimics Lorentz contraction and mass-energy increase[a]. Space-time in GR is similar to stressed matter[5]; there is a close analogy between sound propagation in background hydrodynamic flow and field propagation in curved space-time[6], and so on. Theories advocating preferred frames and ether-like models have been lately proposed in mathematically rigorous contexts: diverse Einstein-ethers[7], MOND[8], quintessence[9,10], scalar-tensor theory[11,12], dark fluid[13], Chameleon scalar field[14,15], TeVeS[16], and the general notion of condensed vacuum states[17] in quantum mechanical (QM) theories of condensed-matter, specifically in super fluid Helium III[18]. With the advent of stringy universe-on-a-membrane models[19,20,21], space being a medium has entered the mainstream: The membrane is the new medium, woven from strings. Naïve substantivalism cannot be the final word; the nature of space-time is relational on principle[22], but we also cannot anymore ignore that GR is likely emergent.

Still relatively new to enter this discussion is the observation that the cosmos undergoes *accelerated* expansion (section 2). How is metric expansion to be understood[23] if kinematics can be understood as being due to *microscopic* dynamics? Global expansion would have to be thought of as the large scale average of locally acting mechanisms. Can this be consistent with the GR description, where global expansion can be pure kinetics acting on initial boundary conditions? In somewhat too simplifying words: Some theories seem to rejuvenate the old ether, but if space grows, where is the hypothetical medium supposed to originate from? Can any mechanism of proliferation be consistent with what is observed globally *and* locally?

---

[a] A moving Burgers screw dislocation in a crystal contracts to $L' = fL$, where $f^2 = 1-(v/c)^2$ and $c$ is the velocity of transverse sound. The energy is the dislocation's potential energy at rest divided by $f$.

In chaotic and especially eternal inflation[24] models, the Hubble constant $H$ takes on different values in different regions and pocket universes[25]. However, if $H$ does arise in such a dynamical way from microscopic QM physics of an inflaton field, it is also over much smaller lengths to be understood as a *local space reproduction rate* (section 3). This view is illustrated by discussing that even exponential inflation in TeV-gravity is slow on the Planck time ($t_\mathrm{P}$) scale, which is the relevant fundamental time scale if microscopic dynamics is involved. We thus support the from dimensional analysis expected limit

$$H < t_\mathrm{P}^{-1} \qquad (1)$$

Whether global expansion is the observed average due to integrated local expansions or not is hidden in a homogeneous world. The cosmos is inhomogeneous on small scales and a space reproduction rate would depend on position. We therefore discuss models that try to deal with in-homogeneity (section 4). The Einstein-Straus vacuole[26,27] has several drawbacks. We set up a Lindquist-Wheeler like lattice model[28] appropriate for a principally everywhere inhomogeneous cosmos that is homogeneous only on large scales. The expanding lattice grows empty space while the energy-momentum density at the lattice sites may stay constant (section 5). This precludes naïve ideas about matter being locally responsible for reproducing space. It demands that a dynamical mechanism would have to mainly reproduce the empty space around energy density distributions, where Weyl curvature dominates while Ricci curvature is absent. The empty space reproduction rate (local Hubble constant) should be proportional to a factor well known from Kepler's third law. The conclusion and outlook will discuss the possible relevance for the ability to model black holes (BH) in emergent gravity scenarios and the size of the vacuum energy density.

## 2. Interpreting Metric Expansion

When considering a homogeneous universe, classical expansion *through* space and GR's Friedmann-Robertson-Walker (FRW) description fit together seamlessly. The way in which a cloud of Newtonian dust expands *through* space is consistent with the global FRW picture of the universe expanding as a whole. This seems paradoxical: In the Newtonian/special relativistic (SR) description, the underlying space stays the same, uninvolved stage. In the GR picture, space expands in the concrete sense that there is more of it than before; this is obvious when considering closed or compactified universes. Space expands globally, while locally it seems to actually nowhere expand. Space-time resolves this "expansion paradox"[22] tentatively: The space of the past is a different region of space-time[b]. It did *not* grow into the larger space of today; it is still in the past! However, any microscopically dynamical concept potentially reintroduces space as being an entity living *in* time. A hypothetical ether medium cannot by some magic only globally have appeared already without having been locally supplied somewhere - or everywhere, but still with a *locally* acting mechanism. Units of space must locally expand, be created, flow in from the sides or "rain" from a higher dimension on the top.

---

[b] The same cannot be said equally for Newtonian space-time, which is rather space in time while the unique spatial direction is never in doubt (e.g. contested by other observers).

Our universe is globally flat (not closed by curvature) and may also not be compactified. In a flat and infinite universe it is difficult to argue that there is more total space later in cosmological time $t_c$. From inside the universe, $t_c$ is determined by observation of the cosmic microwave background (CMB) and average star background. The cosmological principle states that the background is from anywhere in the universe observed to be about the same, changing only with $t_c$. However, if two regions far from each other (in space-time) experience different CMB temperatures $T$, it may only imply that they are at a different time $t_c$. To violate the principle, a model should lead to a background that is not isotropic. Can an everywhere isotropic background be modeled relying on boosts between equally valid reference systems merely traveling *through* space? Hubble flow is defined as the recession velocity

$$v = DH \qquad (2)$$

at some distance $D$ away from the observer and for a given Hubble constant $H$. Can one setup a homogeneous situation where any boosted object asymptotically joins a Hubble flow to finally observe an isotropic CMB? A coordinate change can make the FRW model look like Minkowski space-time, but homogeneity of constant time surfaces is lost[29]. So indeed, metric expansion implies growing volume, i.e. expansion *of* rather than *through* space. This does not necessarily imply that space gets bigger in some absolute sense. Indeed, it can be helpful to imagine that fundamental units become smaller instead.

## 3. The Reproduction Rate Picture

The smallest volumes (SV) that can be meaningfully considered depend on the Planck length $l_P = \sqrt{\hbar G/c^3}$ and occupy $V \approx l_P^3$. The scale factor $a$ can be thought of as describing the size of the universe. We concentrate mostly on the Hubble constant $H = \dot{a}/a$ because $a$ is unobservable. If $\dot{a} = da/dt_c > 0$ stays constant, $H$ still decreases and the observable expansion slows down. The term "accelerated expansion" is therefore somewhat confusing. The fastest known expansion is exponential inflation lasting a short time $\Delta t_c$:

$$a \propto \exp[H \Delta t_c] \qquad (3)$$

The smoothness of the CMB implies an expansion by a factor of more than $10^{26}$, i.e. $H \Delta t_c \geq 60$. If a dim-dimensional volume inflates by doubling $N$ times, the doubling rate is

$$r = N / \Delta t_c = H \dim / \ln 2 \qquad (4)$$

Inflation terminates with the start of reheating at $t_{reheat}$. This must happen early enough to allow an electro-weak (EW) epoch before the EW-symmetry breaks at $t_c \cong 10^{-12} s$. Hence, inflation may last not longer than $\Delta t_c \approx t_{reheat}$. In other words: The number $t_{reheat} / t_P$ of Planck times that any underlying dynamics may take to accomplish each doubling of the number of SV decreases with the length of $t_P$. Gravity theories with large scale extra dimensions[30,31] have attracted a lot of attention from theorists and experimentalists[32,33] because they may solve the hierarchy problem and also imply that black holes (BH) may be produced[34,35] at particle accelerators. In order to resolve the hierarchy problem with TeV-gravity, the fundamental gravitational Newton constant $G_{TeV}$ needs to be $10^{34}$ times larger than the macroscopically observed *G*. This makes the Planck length

$l_{\text{P-TeV}} = \sqrt{\hbar G_{\text{TeV}}/c^3}$ and in turn $t_{\text{P-TeV}} = l_{\text{P-TeV}}/c$ much longer. TeV-gravity inflation could start reheating at $t_{\text{reheat}} \cong \Delta t_c \cong 10^{10} t_{\text{P-TeV}}$, which is at about $10^{-16}$ s. On a logarithmic scale, this still leaves a long electro-weak epoch. One may consequently assume that the duration of typical inflation scenarios is at least about

$$\Delta t_c \cong 10^{10} t_{\text{P}} \qquad (5)$$

According to (4), $\dim = 3$ dimensional space doubles only $N \geq 260$ times during inflation. In other words, an underlying dynamics has $10^8$ Planck times available to accomplish reproducing just one more SV per SV.

We want to consider a comprehensive toy-model to make the reproduction rate be understood as something that arises quite naturally from local microscopic dynamics. One could imagine that the SV multiply similar to bacteria by dividing at a certain rate of reproduction *H*. The population doubles once per generation with the rate (4). One could think of the SV being nurtured by gravitationally repulsive dark energy: especially phantom energy[36] that would eventually pull apart any bound system in a "big rip"[37,38], would be "feeding" the reproduction. Considering universe-on-a-membrane models[19,20,21], one could imagine that the SV are due to the strings that the membrane is made from. The slowness of inflation is the fact that string excitations could travel the string length $\cong l_P$ about $10^8$ times during every generation, i.e. while reproducing just one more unit of about the string's size. An underlying dynamics is not pressed for time as long as a doubling occurs over many Planck times, i.e. the doubling rate is only limited by $r < t_P^{-1}$. This further justifies the limit (1).

The ongoing expansion of eternal inflation does not squeeze pocket universes like a growing culture of bacteria may push the walls of a Petri dish. It is sufficient to consider the SV dividing (not growing). All scales available to measure metric expansion depend on the Plank scale. The Planck length in turn can be thought of as being based on the average length of an SV.

### 4. Modeling In-homogeneity with a Flat Lattice

The describing of exponential inflation via the doubling of fundamental space units illustrated how the global Hubble constant $H$ can be understood as the reproduction rate of *everywhere* (locally) reproducing space. If space grows differently in different regions, $H$ will depend on location. This local "*space reproduction rate*" $H_{(\vec{r})}$ may not coincide with the local *expansion scalar* $\theta = -K$ of Raychaudhuri's equation, which comes from the trace of the external curvature tensor $K_{ij}$. Both concepts make only sense with a choice of a spatial hyper-surface. In other words, the spatial metric $g_{ij}$ introduces the difficulty of having to specify a certain 3+1 split of space-time, i.e. a decision about what time coordinate to single out. The tensor $K_{ij}$ is often based on the time parallel to the world lines of Newtonian dust particles, but this may generally not be close to a cosmic time that could perhaps be derived from the CMB background. $H_{(\vec{r})}$ will be valid locally for small distances $D$ (compare (2)). On large scales however, the Hubble constant would be the average of $H_{(\vec{r})}$ over some volume that contains the distance $D$ between the observer and the object whose recession velocity is observed:

$$\langle H \rangle = \frac{1}{V} \int H_{(\vec{r})} \sqrt{\det(g_{ij})} dV \qquad (6)$$

This should not be expected to be equal to the observed global Hubble constant $H$. The picture of investigating the distances on top of a golf ball with a tape measure that does not bend down into the dimples applies here. This is analogous to what we naturally do if we investigate cosmic length scales employing red shifted light from very far away, as that light has been mostly traveling through vast voids rather than through "the dimples", where it would have been swallowed by stars and dust. In the golf ball analogy: the light observed from far away is that which did not go through many dimples, but through the flat space between the dimples. I.e. $\langle H \rangle \neq H$.

In-homogeneity has been modeled with the Einstein-Strauss vacuole[26] (ESV). While the space outside the ESV is homogeneous and the metric is FRW, inside it is empty and the Schwarzschild solution (SS) applies. There are many problems with the ESV and also the newer variants[39] of it. They are unstable and only spherically symmetric solutions exist. There is also the drawback that interplanetary densities in solar systems are much higher than in the intergalactic medium[c], not much less like in the empty ESV. That is of no concern if expansion is kinematical due to boundary conditions and the space is static and SS around a central body even in the presence of solar wind. But local energy density is a concern if one wants to address locally acting mechanisms.

Consider instead a lattice of masses $M$ at the lattice sites. An expanding lattice in closed universes can maximally contain 600 lattice sites[28]. An infinite Lindquist-Wheeler like lattice solution has not been described in the literature. Nevertheless, we do not aim at advertising yet another unstable solution to the GR field equations. Our starting point is the already confirmed structure, and according to observation, our universe is for all practical purposes flat and infinite. Therefore, consider an infinite lattice of simple cubic (SC) structure under the crystallographic classification. An expanding lattice is taking into account that the universe is actually never homogeneous: there is always some distance scale below which the cosmological fluid, e.g. the dust (not to be confused with hypothetic ether), breaks down into its constituent particles, e.g. galaxy clusters.

---

[c] The interplanetary medium includes dust and solar wind. Its density is about 5 protons per cubic cm in earth's vicinity and decreases with the inverse square of the distance from the sun. The universe has on average only one proton per cubic meter.

In a homogeneous universe ($H_{(\vec{r})} = \langle H \rangle$), it is interpretation whether we blame expansion on the universe prescribing global kinematical expansion as a boundary condition or whether more space is produced everywhere locally. If in-homogeneity is present however, there may be measurable differences. Any effects would be obviously extremely small because recession velocities are very small for distances *D* comparable to the radius of a galaxy. Moreover, whether Hubble flow can result in frame-dragging at all is still controversial and beyond the scope of this article. Hence, one should take the following calculation as an exercise that establishes the flat lattice as a mathematically sound model. The lattice has unit cells of volume $V = a^3 + C$, where the lattice constant *a* will serve also as the cosmological scale factor *a*. *C* takes account of that the volume close to the masses is influenced by curvature (it is what one dimple adds in the golf ball analogy). As the universe expands, the empty space grows ever larger while the metric around the lattice sites stays basically constant. Thus, *C* becomes asymptotically a negligible constant (we will not discuss re-collapsing solutions, as they may well be unphysical). The on large scales homogeneous density $\rho$ depends on the SC lattice structure (a face centered cubic (FCC) lattice would have a different dependence instead):

$$\rho = M/V = M/a^3 \tag{7}$$

This is the globally "seen" density obtained when looking at the universe with a low resolution that does not resolve the curvature near the lattice points ($C$ is not seen). This is analogous to measuring a golf ball with a tape measure that does not bend down into the dimples. In the lattice, one could think of measuring along the geodesics in between the lattice planes. As discussed above, this is what we naturally do if we investigate cosmic length scales employing red shifted light from very far away. Moreover, $C$ becomes negligible as expansion goes on. If metric expansion depends on the local energies and the observed expansion globally is the resulting average $<H_{(x)}>$ on large scales, locally dense spots have even been suggested to contract[40], yet here we only consider a very conservative model. We look at zero pressure models ($P=0$), so that the global Hubble constant is given by the Friedmann equation ($k=0$, $\lambda=0$ and $G=1$ for simplicity) as follows:

$$H^2 = (\dot{a}/a)^2 = (8\pi/3)\rho \tag{8}$$

Consider an observer positioned unit-cell centered at coordinate $r=0$ in between the eight masses $M$ at the cell's corners. She has initially a distance $D_{in} = a/2$ from the six nearest crystal planes. $D_{in}$ is the proper length[d] prescribed by the global FRW metric, i.e. the square of the eigen-time $\tau$ along any world line is given via:

$$(d\tau)^2 = (dt)^2 - a^2\left[(dr)^2 + r^2(d\Omega)^2\right] \tag{9}$$

With $d\Omega = 0$ and $dt = 0$ follows:

$$D = \left| \int_0^D d\tau \right| = \left| \int_0^{r_D} (a) \, dr \right| = r_D a \tag{10}$$

where $r_D$ is the co-moving radial coordinate of the four nearest masses.

Due to symmetry, the observer stays co-moving with the $r = 0$ coordinate in the centre of the cell. Now we consider a light test particle at coordinate $r_d$ in between the observer and one of the masses $M$. The attraction towards that mass $M$ is arbitrarily small with the choice $d_{in} \ll D_{in}$, where $d$ is the proper distance of the test body. The test body is given a velocity $v_{in} = H d_{in}$ so that it is also initially co-moving with its coordinate $r_d$. That Hubble flow velocity $v = Hd = Har_d$ of the coordinate changes over time:

$$H = \frac{2}{3t} \quad \Rightarrow \quad v = \frac{2d_{in}}{3t_{in}} \tag{11}$$

Considering observer and test body to be inside an empty vacuole prescribes that the test body will just keep going with the initial velocity. Starting at $t_{in}$, the universe has doubled ($D = 2 D_{in}$) in linear size at $t_2 := 2/(3\pi \rho_{in})$, as follows from the scale factor:

$$a^3 = (3A/2)^2 t^2 \quad ; \quad A^2 = (8\pi/3) \rho_{in} a_{in}^3 \tag{12}$$

($A$ is a constant due to $\rho = M/a^3 = M/(2D)^3$). Traveling with constant velocity during $\Delta t = t_2 - t_{in}$ results in the distance between observer and test body to grow to:

$$d_2 = \frac{1 + 4\sqrt{2}}{3} d_{in} \cong 2.22 d_{in} \tag{13}$$

---

[d] This is non-trivial, because space-time is measured with light signals, and the expansion of space influences the outcome of measurements.

This is the accepted GR result and it implies that one should *not* think of metric expansion as being homogeneous. A particular model for microscopic dynamics of metric expansion may claim that space expands homogeneously *everywhere* like globally prescribed. In that case, the test particle will stay co-moving and since the universe doubled, that point of view would predict $d_2 = 2d_{in}$ instead. Although space now locally expands, the distance is *smaller* then above, because the initial velocity just lets the test body co-move while the peculiar velocity is zero. Contrary to acceleration expected due to misconceived frame dragging by global expansion, the result is deceleration. This somewhat counter-intuitive result only confirms a derivation that considered tethers[41]. One should however note the difference to what is usually taught in this context. The lore goes mostly somewhat like this: *Global expansion does not drag on gravitationally or otherwise bound systems, i.e. there is no tendency of the expansion to enlarger your living room, firstly because it is a strongly bound system but more importantly because the local expansion is not equal to the global one, i.e. the local metric is Schwarzschild near a celestial body rather than the FRW.* We agree with most of this, except that the very beginning is wrong already. Un-accelerated global expansion, if it were the same also locally as it is globally, would not even drag on the systems anyways; it would rather push on them! This is why the discovered deceleration makes microscopic dynamical concepts promising for attempting to explain the Pioneer anomaly[e] as has been similarly tried before[40].

---

[e] The Pioneer anomaly is seen in radio Doppler and ranging data, yielding information on the velocity and distance of spacecraft. When all known forces are taken into consideration, a very small but unexplained

A contrived enough mechanism underlying metric expansion may mask any measurable differences, but we do not want to embark on defending any model starting from some prejudice about the possibility of short scale corrections to GR. It is not our aim here to discus hypothetical corrections to GR. The aim of this paper is to stay inside orthodox GR and to see how this constrains microscopic dynamical mechanisms. A natural suggestion one may deduce from scaling arguments as follows.

## 5. Global Metric Expansion consistent with Local Expansion

The dependence of the Hubble constant on energy density $H \propto \rho^{1/2}$ (compare (8)) suggests that space might kind of "flow out" of energy density, but it holds $H = \dot{a}/a$, i.e. volume grows with $\dot{V} = 3HV = \sqrt{24\pi MV}$. The growth $\dot{V}$ itself also growths, but the growing adds empty space while the masses $M$ at the lattice sites and their close environments may stay the same, for example if they are solid spheres or BH. This directs attention to the empty space! In a model consistent with observed global expansion, the fraction of empty volume may be much larger than that with matter, so any weirdness close to the masses' surfaces or SS radii can be neglected as a small correction. The same is valid for the space so far away that curvature is influenced much by mass on other lattice sites. With this in mind, consider a volume $V + C$ centered on one of the masses $M$. We are interested in late times, so we neglect $C$. In other words, we solve (6) with $\sqrt{\det(g_{ij})} = 1$ for $H_{(\vec{r})} = H_{(r)}$ with $r$ now centered at the mass $M$:

---

force remains. It appears to cause a constant sunward acceleration of $(8.74 \pm 1.33) \times 10^{-10}$ m/s² for both probes; by coincidence (?) close to the product of the light velocity $c$ and Hubble constant $H$.

$$H_{(r)} = \frac{d}{dV}(V\langle H\rangle) = \langle H\rangle + V\frac{d}{dV}\langle H\rangle \tag{14}$$

Now consider that $\langle H\rangle$ is proportional to $V^{-1/2}$ (see (8)) and it follows $H_{(r)} = \langle H\rangle/2$.

Since $V$ is here the growing sphere $V = (4\pi/3)r^3$, it follows:

$$H_{(r)} = \sqrt{M/(2r^3)} \tag{15}$$

In result, one can reproduce the global expansion by letting space grow around all masses with the local growth proportional to Kepler's $\dot{\phi} = \sqrt{M/r^3}$ (all valid not too close to the masses and at late times). Open and closed ($k = \pm 1$) Friedmann models and Raychaudhuri's equation hint at that external curvature $K$ influences spatial expansion. However, Raychaudhuri's equation says the same as Einstein's equations of course, namely that a small ball of test masses initially mutually at rest will contract, regardless of whether the universe at large expands or contracts! This shows that the extrinsic curvature $K$ cannot describe an expansion of space that is due to local dynamics unaware of the global affairs. An in depth discussion of this is beyond the scope of the present work, but it should be pointed out again that the topic of local versus global expansion is not completely solved by averaging local Raychaudhuri expansion[42].

Ricci curvature $R_{\mu\nu}$ is the only one used in Einstein's field equation, but it is zero in vacuum. It is therefore counter-intuitive that the local curvature in empty space influences the reproduction rate $H_{(r)}$. The Weyl curvature tensor (conformal tensor $C_{\mu\nu\alpha\beta}$) is the traceless component of the Riemann curvature tensor $R_{\mu\nu\alpha\beta}$. With the Ricci part absent, only the Weyl tensor can be responsible. It has not been considered because the Weyl tensor is zero in homogeneous, isotropic space, and the expanding Friedmann models are homogeneous and isotropic. Nevertheless, microscopically, all cosmological fluids are made from particles in what is mostly empty space, i.e. fundamentally, there is always Weyl curvature. It is beyond the scope of the current work to calculate curvature tensors around the SS and review related or even hypothetical QM physics in order to support a factor of $\sqrt{M/r^3}$. Hence, it is reassuring that a factor with these powers of $M$ and $r$ is well known to arise, namely in Kepler's third law of planetary motion. Moreover, it is well known that QM vacuum polarization (virtual particle –antiparticle creation) can violate the strong equivalence principle (SEP). QM corrections to light-cones in empty, curved space-time[43] yield superluminal photons. QM gravity affects photons through Weyl curvature, leading to gravitational birefringence or the "gravity rainbow".

The lattice model has an important piece of self-consistency: If the masses $M$ are also clouds of dust, their expansion can be modeled in the same way internally. At their surface, they would then have the same (and internally homogeneous on scales that do not resolve the clouds particles) $H_{(r)}$ due to their volume and internal density as when calculated from the outside, namely the Hubble flow $v = \langle H \rangle r$ equals the escape velocity. If the sphere is a dust cloud, we can trust equation (15) all the way down to and through the surface of the sphere. Even the dust's (or lattice) SS radius is unproblematic: the cosmic expansion's Hubble flow does indeed exceed light velocity outside of the Hubble radius. Equation (15) may be trusted for the simple model for which it was derived, i.e. for where space is mostly empty, but not at the surface of a solid sphere or a star with pressures and EM-fields, nor at the SS radius of objects other than expanding dust clouds. Indeed, if the lattice sites are BH, the equations suggest a Hubble flow equal to *c* leaving the region, which seems problematic and a discussion of this is also beyond the present work and will be attempted elsewhere.

## 6. Conclusion and Outlook

The global Hubble constant can be reinterpreted as and is formally a space reproduction rate. We illustrated this by showing that even during the fast expansion in exponential inflation, the fundamental processes are slow. Since the global Hubble rate can be understood as the rate of space reproduction, it can be formally related to local dynamical processes. We introduced a mass distribution in form of an expanding flat lattice in order to make the local emergence of the Hubble constant mathematically concrete. A discussion of the scaling inside the expanding lattice showed that it is specifically empty space that would need to reproduce in order to be consistent with astronomical observations. The scaling resulted in a local Hubble constant related to Weyl curvature and proportional to a factor in Kepler's third law.

Our method of using a local reproduction rate $H_{(r)}$ could be employed to fit BH consistently into certain Einstein-ether concepts. Such has been often attempted, for example based on superfluid flow of 3He[44,45] and atomic Bose condensates[46,47], and based on similarity between BH collapse and the instability of atomic Bose condensates to attractive forces[48], see also this recent work[49] and references therein. Especially fluid models' sinks reproduce the locally unspectacular event horizons (EH) as they are known from GR. Nevertheless, with fluid sinks, Hubble flow towards the inside (not curvature) traps objects behind EH and the issue of how fluid disappears is non-trivial. Our considering cosmic expansion showed that space could consistently reproduce proportional to a local Hubble constant. Therefore, space could certainly be (described as if) contracting inside BH in a process that is mathematically consistent and slow on a fundamental time scale. This may help emergent GR concepts gain consistency regardless of whether the nature of actually existing BH is thus addressed.

The following considerations may help to make the hypothetical connection between Weyl curvature and reproduction rate $H$ mathematically more concrete in future work: Ricci curvature measures how much space-time volume $dVdt$ is shrunk or enlarged by energy-momentum density. In empty space, $R_{\mu\nu} = 0$ and $dVdt$ is only deformed (bent). Around a BH for instance, radius is stretched but the time direction's shrinkage is compensating for it. This is relative to the coordinates chosen and it is not obvious which $t$-coordinate most resembles some maybe preferred cosmological time $t_c$. The eigen-time of a free falling observer is near a BH not the obvious choice, especially since Einstein-ether theories are also subject of our exposition. Now let us consider two facts: Firstly, recall that the found $r$-dependence of $H_{(r)}$ needs to be valid only far away from the SS radius; corrections close by would amount to small corrections consistent with all observations. Further away from the center, the time in Kepler's law is parallel to the cosmological time $t_c$ and therefore unproblematic. Secondly, if only vacuum energy density is present, a flat universe will expand exponentially, i.e. $H \sim \rho^{1/2}$ is constant. This means that QM effects are known to result in vacuum reproducing space everywhere at a certain rate $H_{\text{QM-vac}}$. We only know this rate to be constant relative to $t_c$, i.e. un-deformed time. Now let us put these two considerations together: We just saw how the ratio time/space is shrunk in the SS background's Weyl curvature, and so the rate $H_{\text{QM-vac}}$ may be affected similarly. It is immediately clear that if such a relation exists, the vacuum energy density cannot be actually zero. Whether this necessitates a preferred reference system ($t_c$) or is just more easily understood in this way is not subject of the current work. The exciting possibility of deriving the vacuum energy density from its relation to a consistent Hubble law must be attempted without the convenience of assuming $P = 0$ and $\lambda = 0$ for simplicity.

[1] Janssen, M.: Drawing the line between kinematics and dynamics in special relativity. Studies in Hist. and Phil. of Mod. Phys. **40**, 26-52 (2009)

[2] Volovik, G. E.: Superfluid Analogies of Cosmological Phenomena. Phys. Rept. **351**, 195-348 (2001)

[3] Chapline, G.: Quantum Model for Spacetime. Mod. Phys. Lett. A **7**, 1959-1965 (1992)

[4] Frank, C. F.: On the equations of motion of crystal dislocations. Proceedings of the Physical Society of London, A **62**, 131-134 (1945)

[5] Sakharov, A.: Vacuum quantum fluctuations in Curved Space and the Theory of Gravitation. Sov. Phys. Dokl. **12**, 1040-1041 (1968)

[6] Unruh, W. G.: Experimental Black Hole Evaporation? Phys. Rev. Lett. **46**, 1351-1353 (1981)

[7] Bonvin, C., Durrer, R., Ferreira, P. G., Starkman, G., Zlosnik, T. G.: Generalized Einstein-Aether theories and the Solar System. Phys. Rev. D **77**, 024037 (2008)

[8] Milgrom, M.: A modification of the Newtonian dynamics as a possible alternative to the hidden mass hypothesis. Astrophys. J **270**, 365 (1983)

[9] Caldwell, R. R., Dave, R., Steinhardt, P. J.: Cosmological Imprint of an Energy Component with General Equation of State. Phys. Rev. Lett. **80**, 1582–1585 (1998)

[10] Ratra, B., Peebles, P. J. E.: Cosmological consequences of a rolling homogeneous scalar field. Phys. Rev. D **37**, 3406–3427 (1988)

[11] Sanders, R. H.: A stratified framework for scalar-tensor theories of Modified Dynamics. Astrophys. J. **480**, 492-502 (1997)

[12] Sanders, R. H.: Solar System constraints on multi-field theories of modified dynamics. Mon. Not. R. Astron. Soc. **370**, 1519-1528 (2006)

[13] Arbey, A.: Dark Fluid: A complex scalar field to unify dark energy and dark matter. Phys. Rev. D **74**, 043516 (2006)

[14] Khoury, J., Weltman, A.: Chameleon Fields: Awaiting Surprises for Tests of Gravity in Space. Phys. Rev. Lett. **93**, 171104 (2004)

[15] Khoury, J. and Weltman, A.: Chameleon cosmology. Phys. Rev. D **69**, 044026 (2004)

[16] Bekenstein, J. D.: Relativistic gravitation theory for the modified Newtonian dynamics paradigm. Phys. Rev. D **70**, 083509 (2004)

[17] 't Hooft, G.: In Search of the Ultimate Building Blocks. Cambridge Univ. Press, Cambridge (1997)

[18] Volovik, G. E.: The Universe in a Helium Droplet. Int. Series of Monographs on Phys. **107** (2003)

[19] Brax, P., van de Bruck, C., Davis, A. –C.: Brane World Cosmology. Rept. Prog. Phys. **67**, 2183-2232 (2004)

[20] Khoury, J.: A Briefing on the Ekpyrotic/Cyclic Universe. arXiv:astro-ph/0401579 (2004)

[21] Khoury, J., Ovrut, B. A., Steinhardt, P. J., Turok, N.: The Ekpyrotic Universe: Colliding branes and the origin of the hot big bang. Phys Rev D**64**, 123522 (2001)

[22] Vongehr, S.: Supporting abstract relational space-time as fundamental without doctrinism against emergence. arXiv:0912.3069 (2010)

[23] Bonnor, W. B.: Local Dynamics and the Expansion of the Universe. Gen. Rel. and Grav. **32**, 1005-1007 (2000)

[24] Steinhardt, P. J.: "Natural inflation" in The Very Early Universe. Proceedings of the Nuffield Workshop, Cambridge, 21.6 – 9.7.1982, eds: Gibbons, G. W., Hawking, S. W., Siklos, S. T. C. (Cambridge: Cambridge University Press), pp. 251–66 (1983)

[25] Vilenkin, A.: The birth of inflationary universes. Phys. Rev. D **27**, 2848–55 (1983)

[26] Einstein, A., Strauss, E. G.: The Influence of the Expansion of Space on the Gravitation Fields Surrounding the Individual Stars. Rev. Mod. Phys. **17**, 120-124 (1945)

[27] Schucking, E.: Das Schwarzschildsche Linienelement und die Expansion des Weltalls. Z. Phys. **137**, 595-603 (1954)

[28] Lindquist, R. W., Wheeler, J. A.: Dynamics of a Lattice Universe by the Schwarzschild-Cell Method. Rev. Mod. Phys. **29**, 432-443 (1957)

[29] D. N. Page: No superluminal expansion of the universe. arXiv:gr-qc/9303008 (1993)

[30] Arkani-Hamed, N., Dimopoulos, S., Dvali, G.: The hierarchy problem and new dimensions at a millimeter. Phys. Lett. B **429**, 263-272 (1998)

[31] Arkani-Hamed, N., Dimopoulos, S., Dvali, G.: Phenomenology, astrophysics, and cosmology of theories with submillimeter dimensions and TeV scale quantum gravity. Phys. Rev. D **59**, 086004 (1999)

[32] Argyres, P. C., Dimopoulos, S., March-Russell, J.: Black Holes and sub-millimeter dimensions. Phys. Lett. B **441**, 96-104 (1998)

[33] Emparan, R., Horowitz, G. T., Myers, R. C.: Black Holes Radiate Mainly on the Brane. Phys. Rev. Lett. **85**, 499-502 (2000)

[34] Dimopoulos S., Landsberg, G.: Black Holes at the Large Hadron Collider. Phys. Rev. Lett. **87**, 161602 (2001)

[35] Giddings, S. B., Thomas, S. D.: High energy colliders as black hole factories. Phys. Rev. D **65**, 056010 (2002)

[36] Caldwell, R. R.: A Phantom Menace? Phys. Lett. B **545**, 23–29 (2002)

[37] Caldwell, R. R., Kamionkowski, M., Weinberg, N. N.: Phantom Energy and Cosmic Doomsday. Phys. Rev. Lett. **91**, 071301 (2003)

[38] Nojiri, S., Odintsov, S. D., Tsujikawa, S.: Properties of singularities in (phantom) dark energy universe. Phys. Rev. D **71**, 063004 (2005)

[39] Bonnor, W. B.: A generalization of the Einstein-Straus vacuole. Class. Quant. Grav. **17**, 2739-2748 (2000)

[40] Fahr, H. J., Siewert, M.: Imprints from the global cosmological expansion to the local spacetime dynamics. Naturwissenschaften **95**, 419-425 (2008)

[41] Davis, T. M., Lineweaver, C. H., Webb, J. K.: Solutions to the tethered galaxy problem. Am. J. Phys. **71**, 358-364 (2003)

[42] Buchert, T.: On Average Properties of Inhomogeneous Fluids in General Relativity: Dust Cosmologies. Gen. Rel. Gravity **32**, 105-125 (2000)

[43] Drummond I. T., Hathrell, S. J.: QED vacuum polarization in a background gravitational field and its effect on the velocity of photons. Phys. Rev. D **22**, 343-355 (1980)

[44] Jackobson, T. A., Volovik, G. E.: Event horizons and ergoregions in $^3$He. Phys. Rev. D **58**, 064021 (2006)

[45] Volovik, G. E.: Simulation of Panleve-Gullstrand black hole in thin $^3$He-A. JETP Lett. **69**, 705-713 (1999)

[46] Garay, L. J., Anglin, J. R., Cirac, J. I., P. Zoller, P.: Sonic Analog of Gravitational Black Holes in Bose-Einstein Condensates. Phys. Rev. Lett. **85**, 4643-4647 (2000)

[47] Garay, L. J., Anglin, J. R., Cirac, J. I., P. Zoller, P.: Sonic black holes in dilute Bose-Einstein condensates. Phys. Rev. A **63**, 023611 (2001)

[48] Ueda, M., Huang, K.: Fate of a Bose-Einstein condensate with an attractive interaction. Phys. Rev. A **60**, 3317-3320 (1999)

[49] Chapline, G, Hohlfeld, E., Laughlin, R. B., Santiago, D. I.: Quantum Phase Transitions and the Breakdown of Classical General Relativity. Int. J. Mod. Phys. A **18** 3587-3590 (2003)